\documentclass[11pt,a4paper,english]{article}

\usepackage{babel}
\usepackage{amsmath}
\usepackage{amssymb}
\usepackage{amsfonts}
\usepackage{epstopdf}
\usepackage{graphicx}

\newcommand{\ket}[1]{| #1 \rangle }

\begin{document}

\centerline{\LARGE Iterated Stochastic Measurements}
\vskip 1.0 truecm

\centerline{\large 
Michel Bauer$^{\clubsuit}$\footnote{\texttt{michel.bauer@cea.fr}},
Denis Bernard$^{\spadesuit}$\footnote{Member of CNRS; \texttt{denis.bernard@ens.fr}}
and Tristan Benoist$^{\spadesuit}$\footnote{\texttt{tristan.benoist@ens.fr}}
}

%\maketitle
\bigskip

\centerline{$^{\clubsuit}$ Institut de Physique Th\'eorique%
\footnote{CEA/DSM/IPhT, Unit\'e de recherche associ\'ee au CNRS%
} de Saclay, CEA-Saclay, France.} \centerline{$^{\spadesuit}$ Laboratoire
de Physique Th\'eorique de l'ENS,} \centerline{CNRS $\&$ Ecole Normale
Sup\'erieure de Paris, France.} \medskip{}

\begin{abstract}
  We describe a measurement device principle based on discrete iterations of
  Bayesian updating of system state probability distributions. Although purely
  classical by nature, these measurements are accompanied with a
  progressive collapse of the system state probability distribution during each
  complete system measurement. This measurement scheme finds applications in
  analysing repeated non-demolition indirect quantum measurements. We also
  analyse the continuous time limit of these processes, either in the Brownian
  diffusive limit or in the Poissonian jumpy limit. In the quantum mechanical
  framework, this continuous time limit leads to Belavkin equations which
  describe quantum systems under continuous measurements.
\end{abstract}

\tableofcontents

\section{Introduction}
Informal and formal similarities between Bayesian inference \cite{infer} and
quantum mechanics have been noted quite some time ago, see e.g. \cite{caves}.
Bayesian inference may be seen as a way to update trial probability
distributions by taking into account the partial information one has gained on
the system under study. Indirect quantum measurement consists in obtaining
partial information on a quantum system by letting it interact with another
quantum system, called a probe, and performing a direct Von Neumann measurement
on this probe. Iterating the process of system-probe interaction and probe
measurement increases the information on the system because of system-probe
entanglements.

This has been experimentally implemented in electrodynamics in cavities
\cite{lkb_exp}, but also in superconductor circuits \cite{devoret}. As shown by
these experiments, repeating a large number of times (formally, infinitely many
times) indirect non-demolition measurements \cite{non-demo} reproduces
macroscopic direct measurements with collapse of the system quantum wave
function. Each collapse is stochastic and progressive, becoming sharper and
sharper as the number of indirect measurements increases.

Controlling quantum systems \cite{wiseman} by repeating measurements is, in some
way, as old as quantum mechanics, but it has recently been further developed
aiming at quantum state manipulations and quantum information processing
\cite{Devor}. At a theoretical level, the concept of quantum trajectories
\cite{Qtraj1,Qtraj2} emerges from the need to describe quantum jumps and
randomness inherent to repeated measurements. In parallel, studies of open
quantum systems \cite{open} led to the theory of quantum feedback \cite{wise}
and quantum continual measurements \cite{Qcont}. Belavkin equations
\cite{Belavkin} are stochastic non-linear generalizations of the Schr\"odinger
equation adapted to quantum systems under continual measurements.

Contact between experiments of the type described in ref.\cite{lkb_exp} and
classical stochastic processes was made in ref.\cite{qnd_bb}, showing in
particular that the approach to the collapse is controlled by a relevant
relative entropy. The aim of this note is to follow and complement the study of
ref.\cite{qnd_bb}, by, in some way, reversing the logic. We start by forgetting
quantum mechanics for a while and we study a random process obtained by
discretely and randomly updating a system state probability distribution using
Bayes' rules. Iterated stochastic measurements refer to this random recursive
updating. We describe why and how this leads to a stochastic measurement
principle allowing to measure the initial system state probability distribution
but which implements a random collapse of the system state distribution at each
individual complete system measurement. The initial system state distribution is
nevertheless reconstructed by repeating the complete system measurements. 
We point out a connection between De Fenetti's theorem on exchangeable
random variables, see e.g. ref.\cite{hewitt55}, and iterated stochastic measurements.
We also show that these discrete measurement devices admit continuous formulations with
continual updating. There are two limits: a Brownian diffusive limit in which
the random data used to update the system state distribution are coded into
Brownian motions, this case was studied in ref.\cite{BBB12}, and a Poissonian jumpy
limit in which these random data are coded in point processes. The
construction of the continuous time process relies on deforming an a priori
probability measure on the updating data. The key tool is Girsanov's
theorem.  Then we transport these results, in an almost automatic way, to
quantum mechanics, and we show that quantum mechanical systems under repeated
non-demolition indirect measurements admit a continuous time limit described by
Belavkin equations (\ref{belav_diffu},\ref{belav_poiss}).
This completes results proved in ref.\cite{clement} and makes contact
with those described in ref.\cite{Barchielli2009}.

\section{Iterated indirect stochastic measurements}

Let ${\cal S}$ be the system under study and $A$ be a chosen countable set of
system states $\alpha\in A$ that we shall call {\it pointer states}
\footnote{According to the quantum terminology, but the concept of states is
  here more general as it simply refers to a complete list of labels
  characterizing the system behavior.}. The model apparatus is going to measure
the probability distribution $Q_0(\alpha)$, with $\sum_{\alpha}Q_0(\alpha)=1$,
for the system ${\cal S}$ to be in one of the pointer state.

The model apparatus is made of an infinite series of indirect partial
measurements.  Let $I$ denote the set of possible results of one partial
measurements, which we assume to be finite or countable.  For each system
complete measurement, the output datum is thus an infinite sequence of data
$(i_1,i_2,\cdots)$, $i_k\in I$, associated to the series of successive partial
measurements. The output data are random. The probability distribution
$Q_0(\alpha)$ is to be reconstructed from the sequences $(i_1,i_2,\cdots)$.

To be concrete one may keep in mind that the indirect partial measurements arise
from direct measurements on probes which have been coupled to the system. The
model apparatus is then made of an infinite set of in-going probes -- which, for
simplicity, are supposed to be all identical -- passing through the system
${\cal S}$ and interacting with it one after the other. Measurements are done on
the out-going probes. 

Specifications of the model apparatus depend on the chosen set of pointer
states. One of its manufacturing characteristics is a collection of probability
distributions $p(i|\alpha)$, $\sum_i p(i|\alpha)=1$, for the output partial
measurement to be $i\in I$ conditioned on the system ${\cal S}$ be in the state
${\alpha}\in A$. For simplicity, we shall assume a non-degeneracy hypothesis
which amounts to suppose that all probability distributions $p(\cdot|\alpha)$
are distinct, i.e. for any pair of distinct pointer states $\alpha$ and $\beta$
there exists $i\in I$ such that $p(i|\alpha)\not= p(i|\beta)$.

\subsection{Discrete time description} 

In the model apparatus, a {\it complete measurement} is made of an infinite
series of {\it partial measurements} such that each output of these partial
measurements provides a gain of information on the system. Our first aim is to
decipher which informations one is gaining from the $n^{\rm th}$ first partial
measurements. This will allow us to spell out the way the model apparatus is
working as a measurement device.  

\medskip

$\bullet$ {\it Series of partial measurements and specification of the model
  apparatus}.  Suppose that the first partial measurement gives result $i_1\in
I$. Bayes' law then tells us that the probability for the system ${\cal S}$ to
be in the state $\alpha$ conditioned on the first measurement be $i_1$ is
$Q_1(\alpha|i_1)=Q_0(\alpha)p(i_1|\alpha)/\pi_0(i_1)$, with
$\pi_0(i):=\sum_{\alpha} Q_0(\alpha) p(i|\alpha)$, if $Q_0(\alpha)$ is the
initial probability for the system ${\cal S}$ to be in the state $\alpha$ (this
probability is yet unknown but shall be recovered from the series of partial
measurements making a complete measurement).  Let us now ask ourselves what is
the probability to get $i_2$ as second output partial measurement? By the law of
conditioned probabilities, $\pi_1(i_2|i_1)=\sum_{\alpha} p(i_1,i_2|\alpha)\,
Q_0(\alpha)/\pi_0(i_1)$ with $p(i_1,i_2|\alpha)$ the probability to measure
$i_1$ and $i_2$ on the two first partial measurements conditioned on the system
to be in the state $\alpha$. At this point we need to make an assumption: we
assume that the output partial measurements are independent and identically
distributed (i.i.d.) provided that the system ${\cal S}$ is in one of the
pointer state $\alpha\in A$. This translates into the relation
\begin{eqnarray*}
 p(i_1,i_2|\alpha)=p(i_2|\alpha)\, p(i_1|\alpha),
\end{eqnarray*}
which implies that $\pi_1(i_2|i_1)=\sum_{\alpha} p(i_2|\alpha)\,
Q_1(\alpha|i_1)$. That is the probability $\pi_1(i_2|i_1)$ is identical to the
probability to get $i_2$ as output partial measurement assuming that the system
distribution is $Q_1(\alpha|i_1)$.

Hence, as a defining characteristic property of our model apparatus, we assume
that the output of the $n^{\rm th}$ partial measurements is independent of those
of the $(n-1)$-first outputs provided the system ${\cal S}$ is in one of
the pointer state $\alpha\in A$, that is:
\begin{eqnarray}\label{independ}
  p(i_1,\cdots,i_{n-1},i_n|\alpha) = p(i_n|\alpha)\,
  p(i_1,\cdots,i_{n-1}|\alpha) =\prod_{k=1}^n p(i_k|\alpha),\quad \forall\alpha.
\end{eqnarray}
This specifies our model apparatus. This specification is clearly attached
to the chosen set of pointer states.

Conversely, the pointer states associated to this device are those system states
for which the values of the output partial measurements are independent, i.e.
conditioned on the system to be in a pointer state, the output variables $i_1,\,
i_2,\cdots$ are independent and identically distributed. If the system is
initially in a pointer state $\alpha$, that is its probability distribution is
peaked, $Q_0(\cdot)=\delta_{\cdot;\alpha}$, the occurrence frequency $\nu(i)$
of the value $i$ in the output sequence $(i_1,i_2,\cdots)$ is $p(i|\alpha)$. As
we shall see later, one may then identify the pointer states as the system
states for which independent infinite series of partial measurements (i.e.
independent complete measurements) provide identical occurrence frequencies
$\nu(\cdot)$, and this gives a way to calibrate the device and to determine the
conditioned probabilities $p(\cdot|\alpha)$.

If the system is not in a pointer state, its initial distribution $Q_0(\alpha)$
-- to be determined -- is un-peaked. Let $Q_n(\alpha|i_1,\cdots,i_n)$ be the
probability for the system to be in the state $\alpha$ conditioned on the
$n$-first output partial measurements be $i_1,i_2,\cdots,i_n$. From our
hypothesis (\ref{independ}), the probability to get $i$ as the $n^{\rm th}$
output conditioned on the ${(n-1)}^{\rm th}$ first outputs be
$(i_1,\cdots,i_{n-1})$ is
\begin{eqnarray}\label{probpi}
  \pi_{n-1}(i| i_1,\cdots,i_{n-1})=\sum_{\alpha} p(i|\alpha)\,Q_{n-1}(\alpha|
  i_1,\cdots,i_{n-1}). 
\end{eqnarray}
By Bayes' law, the probability for the system to be in the state $\alpha$
conditioned on the $n$-first measurements be $i_1,i_2,\cdots,i_n$ is then
recursively computed by
\begin{eqnarray}\label{recurr}
 Q_n(\alpha| i_1,\cdots,i_n)= \frac{p(i_n|\alpha)\, Q_{n-1}(\alpha|
   i_1,\cdots,i_{n-1})}{\pi_{n-1}(i_n| i_1,\cdots,i_{n-1})}, 
\end{eqnarray}
where $\pi_{n-1}$ is the probability to get $i_n$ as the $n^{\rm th}$ output. To
simplify notations we denote $Q_n(\alpha| i_1,\cdots,i_n)$ by $Q_n(\alpha)$ and
$\pi_{n-1}(i| i_1,\cdots,i_{n-1})$ by $\pi_{n-1}(i)$. Eq.(\ref{recurr}) can be
solved explicitely:
\begin{eqnarray}\label{Qsol}
  Q_n(\alpha)=Q_0(\alpha)\frac{\prod_ip(i|\alpha)^{N_n(i)}}{\sum_\beta
    Q_0(\beta)\, \prod_ip(i|\beta)^{N_n(i)}},
\end{eqnarray}
with $N_n(i)$ the number of times the value $i$ appears in the $n^{\rm th}$
first outputs.  

Let us point out an interesting reformulation of the above conditions on the outputs
of the model apparatus. A sequence $(i_1,i_2,\cdots)$ of random variables is
called exchangeable if the distribution of $(i_1,i_2,\cdots,i_n)$ is the same as
the distribution of $(i_{\sigma_1},i_{\sigma_2},\cdots,i_{\sigma_n})$ for each
$n$ and each permutation $\sigma$ of $[1,2,\cdots,n]$. A remarkable theorem due
to De Finetti (see e.g. ref.\cite{hewitt55} or the last two items of ref.\cite{proba}) asserts that an
infinite sequence $(i_1,i_2,\cdots)$ of random variables is exchangeable if and
only if there is a random variable $A$ such that, conditionally on $A$,
$(i_1,i_2,\cdots)$ is a sequence of independent identically distributed random
variables. In our construction, the values taken by $A$ are nothing but the
pointer states and the measure on $A$ is $Q_0$. So the hypotheses on the model
apparatus can be rephrased as the fact that the order of partial measurements is
immaterial.

\medskip

More concretely, let $\Omega$ be the data set of all complete measurements. This is
made of all infinite series $\omega:=(i_1,i_2,\cdots)$, $i_k\in I$, of output
partial measurements. We may endow  $\Omega$  with the filtration $\mathcal{F}_n$ of
$\sigma$-algebras generated by the sets
$B_{i_1,\cdots,i_n}:=\{\omega=(i_1,\cdots,i_n,{\rm anything\ else})\in
\Omega\}$, i.e. $\mathcal{F}_n$ codes for the knowledge of the $n^{\rm th}$
first partial measurements. This filtered space is equipped with a probability
measure recursively defined by
$\mathbb{P}[i_n=i|\mathcal{F}_{n-1}]=\pi_{n-1}(i)$. Notice that, given
$Q_0(\alpha)$, this probability measure decomposes as a sum
\[
\mathbb{P}=\sum_\alpha Q_0(\alpha)\, \mathbb{P}_\alpha
\]
where $\mathbb{P}_\alpha$ will be the probability measure induced on $\Omega$ if
the system happened to be initially in the pointer state $\alpha$, i.e. if
$Q_0(\cdot)$ is peaked at $\alpha$. Under $\mathbb{P}_\alpha$ the partial
outputs are independent random variables so that
\[ 
\mathbb{P}_\alpha[B_{i_1,\cdots,i_n}]= \prod_{k=1}^n p(i_k|\alpha).
\]

Let us then quote properties of the random probability distribution
$Q_n(\cdot)$, which will be keys for specifying the model measurement device:

\medskip

\indent (i) Peaked distributions are stable under the recursion relation
(\ref{recurr}). That is, if $Q_0(\cdot)=\delta_{\cdot;\alpha}$ then
$Q_n(\cdot)=\delta_{\cdot;\alpha}$ for any $n$.\\
\indent (ii) Given $Q_0(\cdot)$ generic, the random variables $Q_n(\alpha)$
converge as $n$ goes to infinity almost surely and in $\mathbb{L}^1$. The
limiting distribution $Q_\infty(\cdot)$ is peaked at a random target pointer
state. That is:
\[ Q_\infty(\cdot)=\delta_{\cdot;\gamma_\omega}\] with target pointer state
$\gamma_\omega$ depending on the event $\omega$. The probability for the target
to be a given pointer state $\alpha$ is the initial probability distribution:
\[\mathbb{P}[\gamma_\omega=\alpha]=Q_0(\alpha).\]
\indent (iii) The asymptotic occurrence frequencies $\nu(i):=\lim_n N_n(i)/n$,
with $N_n(i)$ the number of times the value $i$ appears in the $n^{\rm th}$
first outputs, are those of the target pointer state. That is:
\[\lim_{n\to\infty} N_n(i)/n=p(i|\gamma_\omega).\]
\indent (iv) The convergence is exponentially fast: 
\[Q_n(\alpha)\simeq \exp(-nS(\gamma_\omega|\alpha)), \quad
\alpha\not=\gamma_\omega,\] 
for $n$ large enough, with $S(\gamma_\omega|\alpha)$ the relative entropy of
$p(\cdot|\gamma_\omega)$ relative to $p(\cdot|\alpha)$.  

\medskip

These facts have been proved in ref.\cite{qnd_bb}. They are based on the fact that
the random variables $Q_n(\alpha)$ are bounded $\mathbb{P}$-martingales with
respect to the filtration $\mathcal{F}_n$. That is
$\mathbb{E}[Q_n(\alpha)|\mathcal{F}_{n-1}]=Q_{n-1}(\alpha)$. A classical theorem
of probability theory \cite{proba} says that a bounded martingale converges
almost surely and in $\mathbb{L}^1$, so that $Q_\infty(\alpha):=\lim_n
Q_n(\alpha)$ exists and
$Q_n(\alpha)=\mathbb{E}[Q_\infty(\alpha)|\mathcal{F}_n]$. More general results,
involving for instance extra randomness on the partial measurements or relaxing
the non-degeneracy hypothesis on the conditioned probability $p(\cdot|\alpha)$,
have been obtained in ref.\cite{BBB12}.

\medskip

\begin{figure}
\mbox{\includegraphics[width=\textwidth]{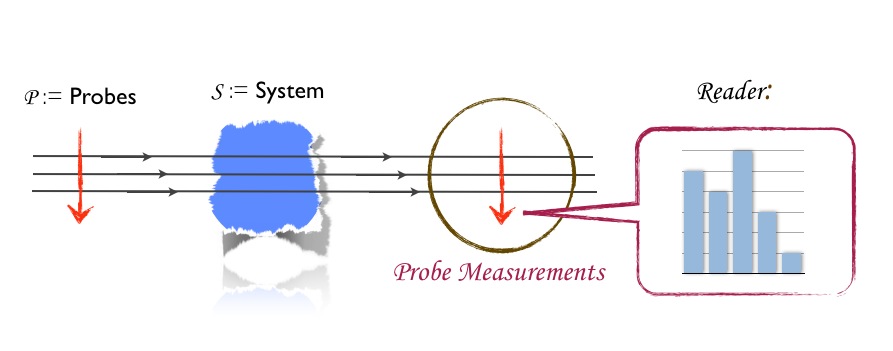}}
\caption{\emph{A schematic view of iterated stochastic measurements: probes are
    send one after the other to interact with the system for a while. After the
    interaction, a measurement is performed on each probe. The information
    gained is summarized in the occurrence frequencies, which allow to identify
    the limiting state.}}
\label{fig:mesure}
\end{figure}

$\bullet$ {\it How to read-off a complete measurement and consequences.}
Let us summarize how the model apparatus is (concretely) working and how data
are analysed, see Fig.\ref{fig:mesure}. For a given system measurement, the data is an infinite sequence $\omega=(i_1,i_2,\cdots)$ of output partial measurements. From its asymptotic behaviour, the apparatus computes the asymptotic frequencies $\nu(i)$ of occurrences of the values $i$ in the sequence $\omega$, and it compares it to one of the apparatus data-base distributions $p(i|\alpha)$. By the non-degeneracy hypothesis and the above convergence theorem \cite{qnd_bb}, each of the asymptotic frequencies coincide with one of the data-base distributions, so that the comparison identifies uniquely the target pointer state and that identified state is by definition the result of a complete system measurement. Since by the above theorem the distribution of the target pointer states is the initial distribution $Q_0(\cdot)$, the histogram of repeated independent complete system measurements yields the initial distribution.

Notice that by the end of a complete measurement the system state distribution has collapsed into one of the pointer states. The need for an infinite series of partial measurement reflects the need for a macroscopic apparatus to implement the collapse. If the system measurement is stopped after a finite number of partial measurements the collapse is only partial, i.e. the probability distribution $Q_n(\cdot)$ is still smeared around the target pointer state. The target pointer state may nevertheless be identified with high fidelity if the differences between the data-base probability distributions $p(\cdot|\alpha)$ are bigger than the fluctuations of the frequencies $\nu_n(\cdot)$ which generically scale like $n^{-1/2}$.

\subsection{Continuous time limit}

We now describe continuous time limits of the previous model apparatus in which
the partial measurements are done continuously in time. There are different
continuous time limits, depending on the behaviour of the data-base conditioned
probability distributions $p(\cdot|\alpha)$: a Brownian diffusive limit, a
Poissonian jumpy limit, or a mixture of them.

These limits may be understood by looking at properties of the counting process
$N_n(i):=\sum_{k=1}^n \mathbb{I}_{i_k=i}$ which is the number of times the value
$i$ appears in the $n^{\rm th}$ first partial measurements. Recall that
$\pi_{m-1}(i)=\mathbb{E}[\mathbb{I}_{i_m=i}|\mathcal{F}_{m-1}]$ is the
probability to get $i$ as the $m^{\rm th}$ partial output conditioned on the
$(m-1)^{\rm th}$ first partial outputs. We may tautologically decompose $N_n(i)$
as
\begin{eqnarray} \label{Ndoob} 
N_n(i) = X_n(i) + A_n(i),\quad {\rm with}\ A_n(i):=\sum_{m=0}^{n-1} \pi_m(i),
\end{eqnarray}
where this equation serves as the definition of $X_n(i)$, i.e.
$X_n(i):=N_n(i)-A_n(i)$, with $\sum_i X_n(i)=0$ as both $N_n(i)$ and $A_n(i)$
add up to $n$. Then by construction
$\mathbb{E}[X_n(i)|\mathcal{F}_{n-1}]=X_{n-1}(i)$, so that the processes
$X_n(i)$ are $\mathbb{P}$-martingales with respect to the filtration
$\mathcal{F}_n$. Equation (\ref{Ndoob}) is the so-called Doob decomposition of
$N_n(i)$ as the processes $A_n(i)$ are predictable, i.e. $A_n(i)$ is
$\mathcal{F}_{n-1}$-mesurable, see ref.\cite{proba}. The martingale property in
particular implies that $\mathbb{E}[X_n(i)]=0$.

Recall now the recursion relation (\ref{recurr}) that we may rewrite as
\[ Q_n(\alpha)-Q_{n-1}(\alpha) = Q_{n-1}(\alpha) \sum_i
\frac{p(i|\alpha)}{\pi_{n-1}(i)}\, ( \mathbb{I}_{i_n=i}- \pi_{n-1}(i) ) \] which
holds true because $\sum_ip(i|\alpha)=1$. By construction $\mathbb{I}_{i_n=i}-
\pi_{n-1}(i)= X_n(i)-X_{n-1}(i)$, so that
\begin{eqnarray}\label{discrete}
  (\Delta Q)_n(\alpha) = Q_{n-1}(\alpha) \sum_i
  \frac{p(i|\alpha)}{\pi_{n-1}(i)}\, (\Delta X)_n(i).
\end{eqnarray}
with $(\Delta Q)_n(\alpha):=Q_n(\alpha)-Q_{n-1}(\alpha)$ and $(\Delta X)_n(i):=X_n(i)-X_{n-1}(i)$.
We thus have rewritten the recursion relation (\ref{recurr}) as a discrete
non-linear difference equation for the probability distribution $Q_n(\cdot)$
driven by discrete differences of the martingales $X_n(i)$. This will be the
starting point of the continuous time limits.  \medskip

Before going on let us point out a geometrical interpretation of $Q_n(\alpha)$ which
will be useful later. On the set of complete measurements, we have defined a
global measure $\mathbb{P}$ and a series of measures $\mathbb{P}_\alpha$
associated to each of the pointer states with $\mathbb{P}=\sum_\alpha
Q_0(\alpha) \mathbb{P}_\alpha$. It is a simple matter to check that
$\mathbb{P}_\alpha$ is non singular with respect to $\mathbb{P}$, so that there
exists a Radon-Nikodym derivative of $\mathbb{P}_\alpha$ with respect to
$\mathbb{P}$, see ref.\cite{proba}. This derivative is
$Q_\infty(\alpha)/Q_0(\alpha)$. More concretely, for any
$\mathcal{F}_n$-measurable integrable function $X$, 
\[ Q_0(\alpha)\, \mathbb{E}_\alpha[X]= \mathbb{E}[Q_n(\alpha)\,X],\] with
$Q_n(\alpha)=\mathbb{E}[Q_\infty(\alpha)|\mathcal{F}_n]$, as may be checked
directly.

We may tautologically refine this geometrical construction. Let us start from an
arbitrary probability measure $\mathbb{P}^0$ on $\Omega$, and let us set
$\mathfrak{z}_{i_1,\cdots,i_n}:= \mathbb{P}^0[B_{i_1,\cdots,i_n}]$, assuming
that none of these probabilities vanish. Let $Z_n$ and $Z_n(\alpha)$ be
$\mathcal{F}_n$-measurable functions defined by
\[ Z_n(\alpha)(i_1,\cdots,i_n):=\mathfrak{Z}^{-1}_{i_1,\cdots,i_n}\, \prod_k
p(i_k|\alpha),\quad Z_n:=\sum_\alpha Q_0(\alpha)Z_n(\alpha),\] so that
$Q_n(\alpha)/Q_0(\alpha)=Z_n(\alpha)/Z_n$. It is easy to check that each
$Z_n(\alpha)$ is a $\mathbb{P}^0$-martingale,
$\mathbb{E}^0[Z_n(\alpha)|\mathcal{F}_{n-1}]=Z_{n-1}(\alpha)$. As it is clear
from their definition, the $Z_n(\alpha)$'s are the Radon-Nikodym derivative of
the measures $\mathbb{P}_\alpha$'s with respect to $\mathbb{P}^0$ on
$\mathcal{F}_n$-measurable functions, that is
\[ \mathbb{E}_\alpha[X]= \mathbb{E}^0[Z_n(\alpha)\,X],\]
for any $\mathcal{F}_n$-measurable function $X$. Of course, $Z_n$ is the
Radon-Nikodym derivatives of $\mathbb{P}$ with respect to $\mathbb{P}^0$, i.e.
$\mathbb{E}[X]= \mathbb{E}^0[Z_n\,X]$ for any $\mathcal{F}_n$-measurable
function $X$. Choosing adequately $\mathbb{P}^0$ helps taking the continuous
time limit, a fact that we shall illustrate below.

\subsubsection{Brownian diffusive limit}
The Brownian diffusive limit occurs when the conditioned probability $p(\cdot|\alpha)$ depends on an extra small parameter $\delta$ such that 
\[ p(i|\alpha)\simeq_{\delta\to0} p_0(i)\, \big( 1+\sqrt{\delta}\,
\Gamma(i|\alpha)+\cdots \big),\] with all $p_0(i)$'s non vanishing and
$\alpha$-independent. Since $\sum_ip_0(i)=1$, the $p_0(\cdot)$'s define an
$\alpha$-independent probability measure on $I$. 
Note that $\sum_i p_0(i)\Gamma(i|\alpha)=0$ for all $\alpha$ since $\sum_ip(i|\alpha)=1$ for all $\delta$. 
By the non-degeneracy assumption, the functions $\Gamma(\cdot|\alpha)$ on $I$ are all different.

The continuous time limit is then obtained by performing the scaling limit $\delta\to0$, $n\to\infty$ with $t:=n/\delta$ fixed.

To understand the scaling limit of the counting processes $N_n(i)$, let us look
at its behaviour under $\mathbb{P}_\alpha$, i.e. for a system in the pointer
state $\alpha$ with initial distribution $Q_0(\cdot)=\delta_{\cdot;\alpha}$.
Then by hypothesis the output partial measurements are random independent
variables, so that $N_n(i)=\sum_{k=1}^n \mathbb{I}_{i_k=i}$ is the sum of $n$
independent identically distributed (i.i.d.) variables $\epsilon_k(i)$ with
value $1$ (if the output of $k^{\rm th}$ partial measurement is $i$) with
probability $p(i|\alpha)$ and zero (if the output of $k^{\rm th}$ partial
measurement is different from $i$) with complementary probability. By the law of
large numbers, the $N_n(i)$'s at large $n$ become Gaussian processes with mean
$n\,p(i|\alpha)$ and covariance ${\rm
  min}(n,m)\,(p(i|\alpha)\delta_{i;j}-p(i|\alpha)p(j|\alpha))$. Under these
hypotheses, the probability $\pi_{m-1}(i)$ for the $m^{\rm th}$ output partial
measurement to be $i$ is $p(i|\alpha)$ for all $m$, so that $A_n(i)=n\,
p(i|\alpha)$. Hence, under $\mathbb{P}_\alpha$ and for such peaked initial
distribution, the law of the processes $X_n(i)$ at large $n$ is that of Gaussian processes with zero mean and
covariance
\[
\mathbb{E}_\alpha[ X_n(i)X_m(j)]= {\rm min}(n,m)\,(p(i|\alpha)\delta_{i;j}-p(i|\alpha)p(j|\alpha)),
\quad {\rm for}\ Q_0(\cdot)=\delta_{\cdot;\alpha}.
\]
After appropriate rescaling, this clearly admits a finite limit as $\delta\to 0$
which is $\alpha$-independent. Hence under this hypothesis, $X_n(i)$ admits a
continuous time limit $X_t(i)$, formally to be thought of as
$\lim_{\delta\to0}\sqrt{\delta} X_{[t/\delta]}(i)$. However, the previous
equation is not enough to describe this limit under the law $\mathbb{P}$ and
some care has to be taken, see ref.\cite{BBB12}.

So, let us define the scaling diffusive limit of the state distribution and the
Doob martingales:
\[ Q_t(\delta):=\lim_{\delta\to0} Q_{[t/\delta]}(\alpha),\quad X_t(i):=
\lim_{\delta\to0}\sqrt{\delta} X_{[t/\delta]}(i),\] 
and of the counting processes,
\[ W_t(i):=\lim_{\delta\to0}\sqrt{\delta} (N_{[t/\delta]}(i)-p_0(i)t/\delta).\]
These equalities have to be thought in law, but we shall still denote by
$\mathbb{P}=\sum_\alpha Q_0(\alpha)\mathbb{P}_\alpha$ the probability measure
for the continuous time processes. By construction, $X_t(i)$ are
$\mathbb{P}$-martingales.

The discrete difference equation (\ref{discrete}) naively translates into the
non-linear stochastic equation for $Q_t(\alpha)$. Recall that in the diffusive limit,  $p(i|\alpha)\simeq p_0(i)[1+\sqrt{\delta}\,
\Gamma(i|\alpha)+\cdots]$ as $\delta$ goes to zero, so that $\pi_{n-1}(i)\simeq p_0(i)[1+\sqrt{\delta}\,\langle\Gamma(i)\rangle_t+\cdots]$ with 
$\langle\Gamma(i)\rangle_t:=\sum_\alpha \Gamma(i|\alpha)\, Q_t(\alpha)$. In the continuous time limit, eq.(\ref{discrete}) then becomes:
\begin{eqnarray}\label{conti}
  dQ_t(\alpha)= Q_t(\alpha) \sum_i \big( \Gamma(i|\alpha)-\langle\Gamma(i)\rangle_t \big) dX_t(i)
 \end{eqnarray}
 with It\^o's convention. We used $\sum_i X_t(i)=0$ and $\sum_i p_0(i)\Gamma(i|\alpha)=0$ to take this limit. 
 Remark that this equation preserves the normalisation condition
$\sum_\alpha Q_t(\alpha)=1$. This equation is that which governs
 the evolution of the system probability distribution under continuous Bayes'
 updating in the diffusive limit. The random fields $X_t(i)$
 code for the information of the continuous time series of partial measurements.
 Not all of these fields are independent since $\sum_iX_t(i)=0$.
 As we shall see, the main feature of the Brownian diffusive limit is that the $X_t(i)$'s are
 Gaussian processes with zero mean and covariance
\begin{eqnarray}\label{brown}
\mathbb{E}[ X_t(i)X_s(j)]= {\rm min}(t,s)\,(p_0(i)\delta_{i;j}-p_0(i)p_0(j)),
\end{eqnarray}
Alternatively, the fields $X_t(i)$ are zero mean Gaussian martingales with quadratic variation
\[ dX_t(i)dX_t(j)=  (p_0(i)\delta_{i;j}-p_0(i)p_0(j))\, dt,\]
which is of course compatible with the relation $\sum_iX_t(i)=0$.
Actually the proofs of equation (\ref{conti}) and of the
correctness of (\ref{brown}) are a bit tricky, see \cite{BBB12} for details. We
shall here present an alternative less rigorous but quicker and simpler
argument.  
\medskip

Let us now argue for eq.(\ref{brown}). Recall the Doob decomposition of the
counting process $N_n(i)=X_n(i)+A_n(i)$. Its naive scaling limit reads
\[ W_t(i) = X_t(i) + \int_0^tds\,  p_0(i)\, \langle\Gamma(i)\rangle_s,\]
whose infinitesimal differential form is
\begin{eqnarray}\label{counts}
dW_t(i)=dX_t(i)+ p_0(i)\, \langle\Gamma(i)\rangle_t\, dt
\end{eqnarray}

Contrary to the $X_t(i)$'s, the $W_t(i)$'s are not $\mathbb{P}$-martingales but
they are globally defined and independent of the initial distribution
$Q_0(\cdot)$ because they are defined as limit of the counting process. We
however know that, under $\mathbb{P}_\alpha$, the $W_t(i)$'s are Gaussian
processes with mean and covariance
\begin{eqnarray*} 
\mathbb{E}_\alpha[W_t(i)]&=&tp_0(i)\Gamma(i|\alpha),\\
\mathbb{C}{\rm ov}_\alpha[W_t(i)W_s(j)]&=&{\rm min}(t,s)\,(p_0(i)\delta_{i;j}-p_0(i)p_0(j)).
\end{eqnarray*} 
  We now would like to use this
information to read off properties of the martingales $X_t(i)$'s. The key point
consists in using Girsanov's theorem \cite{proba}. Recall that
$Q_\infty(\alpha)$ may be thought as the Radon-Nikodym derivative of
$\mathbb{P}_\alpha$ with respect to $\mathbb{P}$ and that
$Q_t(\alpha)=\mathbb{E}[Q_\infty(\alpha)|\mathcal{F}_t]$. Assume for a while
that the $X_t(i)$'s are Gaussian processes under $\mathbb{P}$ with zero mean and
quadratic variation $G(i,j)dt:=\langle dX_t(i),dX_t(j)\rangle$ to be determined.
Girsanov's theorem tells us that modifying the measure $\mathbb{P}$ by
multiplication by the martingale $Q_t(\alpha)$ adds a supplementary drift in the
stochastic differential equation ({\ref{counts}), given by the logarithmic
  derivative of the martingales. In the present case, using eq.(\ref{conti}) Girsanov's theorem implies that
\[ dW_t(i)= d\hat X_t(i) + p_0(i)\, \langle\Gamma(i)\rangle_t\, dt + \sum_j
G(i,j)\big( \Gamma(j|\alpha)-\langle\Gamma(j)\rangle_t \big)dt,\] 
with $\hat X_t(i)$'s Gaussian processes under $\mathbb{P}_\alpha$ with zero mean
and identical quadratic variation $G(i,j)dt$. Comparing now with the known
properties of $W_t(i)$ under $\mathbb{P}_\alpha$, spelled out above, we deduce
that $G(i,j)=(p_0(i)\delta_{i;j}-p_0(i)p_0(j))$, as claimed~\footnote{This expression for the quadratic variation of the $X_t(i)$'s is compatible with the relation $\sum_iX_t(i)=0$, since $\sum_i G(i,j)=0$ as it should.}, so that the previous
equation reduces to
\[dW_t(i)= d\hat X_t(i)+p_0(i)\Gamma(i|\alpha)dt,\] 
under $\mathbb{P}_\alpha,$ as required. This ends our argument for equations (\ref{conti},\ref{brown}). 

A way to rigorously construct processes with all the above properties is to deform a suitable a priori measure $\mathbb{P}^0$. Details have been given in ref.\cite{BBB12}. In this note, we shall illustrate this strategy in the Poissonian case. 

Eq.(\ref{conti}) may actually be integrated explicitly, see ref.\cite{BBB12}. Furthermore, as bounded martingales the $Q_t(\alpha)$'s again converge almost surely and in $\mathbb{L}^1$. Under the non-degeneracy assumption that all $\Gamma(\cdot|\alpha)$ are different, the limit distribution is peaked, $Q_\infty(\cdot)=\delta_{\cdot;\gamma_\omega}$, at a random target pointer state. The convergence is still exponential. 

\subsubsection{Poissonian jumpy limit}
The Poissonian limit occurs when the conditioned probabilities $p(i|\alpha)$
vanish as a small parameter $\delta$ vanishes. Not all $p(i|\alpha)$'s may
vanish simultaneously as they sum up to $1$. So let us single out one value
$i^*$ for which $p(i|\alpha)$ goes to $1$ as $\delta\to0$ for all $\alpha$ and
assume that all other $p(i|\alpha)$ vanish in this limit:
\[ p(i^*|\alpha)\simeq_{\delta\to0}1,\quad p(i|\alpha)\simeq_{\delta\to0}
\delta\, \theta(i|\alpha)\quad {\rm for}\ i\not=i^*,\ \forall \alpha. \] 
By consistency
$p(i^*|\alpha)=1-\delta(\sum_{i\not=i^*}\theta(i|\alpha))+o(\delta)$ and all
$\theta(i|\alpha)$ are positive and assumed to be non-vanishing. In the limit
$\delta\to0$, the output of the partial measurements is most frequently $i^*$
with sporadic jumps to another value $i$ different from $i^*$~\footnote{We may
  generalize this by assuming that more than one conditioned probabilities
  remain finite as $\delta$ goes to zero. In these cases, the continuous time
  limit will be a mixture between the Brownian and Poissonian limits.}.

The continuous time limit is obtained by performing the scaling limit
$\delta\to0$, $n\to\infty$ with $t=n/\delta$ fixed. 

To understand the continuous time limit of the counting processes $N_n(i)$, let
us again look at its behaviour under $\mathbb{P}_\alpha$. As before, the output
partial measurements are then random independent variables, so that
$N_n(i)=\sum_{k=1}^n \mathbb{I}_{i_k=i}$ is the sum of $n$ independent
identically distributed variables $\epsilon_k(i)$ with value $1$ with
probability $p(i|\alpha)$ and zero with complementary probability. Let us first
consider $i\not=i^*$ and compute $\log
\mathbb{E}_\alpha[e^{zN_n(i)}]=n\log[(1-p(i|\alpha)+e^zp(i|\alpha)]$. In the
scaling limit with $n=t/\delta$ and $p(i|\alpha)\simeq\delta\, \theta(i|\alpha)$ we
get
\[ \lim_{\delta\to0}\log \mathbb{E}_\alpha[e^{zN_{[t/\delta]}(i)}]= t
\theta(i|\alpha)\, (e^z-1),\quad i\not=i^*.\] 
Similar computations, based on the general formula
\[
{\mathbb{E}_\alpha}\left[e^{\sum_{l=1}^{k}\sum_{i}z_{l}(i)(N_{n_{l}}(i)-
    N_{n_{l-1}}(i))}\right]= \prod_{l=1}^{k}\left(\sum_{i}e^{z_{l}(i)}
  p(i|\alpha)\right)^{n_{l}-n_{l-1}}, 
\]
for $k\geq1$, arbitrary non-decreasing sequences of integers $0=n_{0}\leq
n_{1}\leq\cdots\leq n_{k}$ of length $k$, and arbitrary (complex) $z_{l}(i)$'s,
show that, in the limit $\delta\to 0$, the $\mathbb{P}_\alpha$-distributions of
the counting processes $N_{[t/\delta]}(i)$, $i\not=i^*$, converge to those of
independent Poisson point processes with intensities $\theta(i|\alpha)\,dt$.
Note that this statement is true under $\mathbb{P}_\alpha$ but not under
$\mathbb{P}$. However, we can compute their $\mathbb{P}$-generating functions
using the decomposition of the measure $\mathbb{P}=\sum_\alpha
Q_0(\alpha)\mathbb{P}_\alpha$. For instance
\[
\mathbb{E}[e^{zN_{[t/\delta]}(i)}]\simeq_{\delta\to 0} \sum_\alpha Q_0(\alpha)
e^{t \theta(i|\alpha)\, (e^z-1)},\quad i\not=i^*.\] 

The properties of $N_n(i^*)$ and their limits are reconstructed using the sum
rule, $\sum_i N_n(i)=n$. In particular, for small $\delta$,
$N_{[t/\delta]}(i^*)\simeq t/\delta$ up to order $1$ random corrections.

So, let us define the scaling Poisson limits of the state distribution and of
the Doob martingales $X_n(i)$'s, 
\[Q_t(\alpha):=\lim_{\delta\to0} Q_{[t/\delta]}(\alpha),\quad Y_t(i):=
\lim_{\delta\to0} X_{[t/\delta]}(i)\]  
and of the jump counting processes
\[N_t(i):=\lim_{\delta\to0} N_{[t/\delta]}(i),\quad {\rm for}\ i\not=i^*,\] 
and $M_t(i^*):= \lim_{\delta\to0} \big( N_{[t/\delta]}(i^*)-t/\delta\big)$.
Again, these equalities have to be thought in law, but we still denote by
$\mathbb{P}=\sum_\alpha Q_0(\alpha)\mathbb{P}_\alpha$ the probability measure
for the time continuous processes. By construction, the martingales $Y_t(i)$ sum
up to zero, $\sum_i Y_t(i)=0$, and have zero mean, $\mathbb{E}[Y_t(i)]=0$.
Similarly, $M_t(i^*)+\sum_{i\not=i^*}N_t(i)=0$.

Again, the naive scaling limit of the difference equation (\ref{discrete}) yields a stochastic equation 
for the system state distribution. In the Poissonian limit, one has $p(i|\alpha)\simeq\delta\theta(i|\alpha)+\cdots$ for $i\not= i^*$ as $\delta\to 0$, so that $\pi_{n-1}(i)\simeq \delta\, \langle \theta(i)\rangle_t+\cdots$ with $\langle \theta(i)\rangle_t:=\sum_\alpha \theta(i|\alpha)Q_t(\alpha)$, for $i\not= i^*$, whereas both $p(i^*|\alpha)$ and $\pi_{n-1}(i^*)$ approach $1$ as $\delta$ goes to zero. The continuous time limit of eq.(\ref{discrete}) is then
\begin{eqnarray} \label{poiss-sto}
dQ_t(\alpha) = Q_t(\alpha) \sum_{i\not=i^*} \Big(
\frac{\theta(i|\alpha)}{\langle \theta(i)\rangle_t}-1\Big)\, dY_t(i) . 
\end{eqnarray}
where we used $dY_t(i^*)=-\sum_{i\not= i^*}dY_t(i)$ to deal with the term associated to $i^*$ in eq.(\ref{discrete}).
 As we shall show just below, the $Y_t(i)$'s are related to the counting processes by 
\begin{eqnarray}\label{dpoisson} 
dN_t(i)= dY_t(i) + \langle \theta(i)\rangle_t\,dt, \quad i\not=i^*,
\end{eqnarray}
We shall furthermore argue that the processes $dN_t(i)$, $i\not=i^*$, are point
processes with intensities $\langle \theta(i)\rangle_t\,dt$. This intensity is
sample dependent -- a point that we shall explain --, but predictable. Equations
(\ref{poiss-sto},\ref{dpoisson}) are those which governs the evolution of the
system probability distribution under continuous Bayes' updating in the
Poissonian limit. The random counting processes $N_t(i)$ code for the
informations on the continuous time series of partial measurements.  \medskip

Let us now argue for eq.(\ref{dpoisson}). Consider again the Doob decomposition
$N_n(i)=X_n(i)+A_n(i)$. Because $\pi_{[t/\delta]}(i)\simeq\delta\, \langle
\theta(i)\rangle_t$ for small $\delta$, its naive scaling reads
\[ N_t(i)= Y_t(i) + \int_0^tds\, \langle \theta(i)\rangle_s,\quad i\not=i^*.\]
Its infinitesimal version is eq.(\ref{dpoisson}), as announced. Since
$p(i^*|\alpha)\simeq 1-\delta\sigma(i^*|\alpha)$ with
$\sigma(i^*|\alpha):=\sum_{i\not=i^*}\theta(i|\alpha)$, the counting function
$N_n(i^*)$ slightly deviates from $n$, and $M_t(i^*)=Y_t(i^*) - \int_0^tds\,
\langle \sigma(i^*)\rangle_s$ with $\langle \sigma(i^*)\rangle_s:=\sum_\alpha
\sigma(i^*|\alpha)Q_s(\alpha)$.

By the martingale property, $\mathbb{E}[dY_t(i)|\mathcal{F}_t]=0$ so that 
\[ \mathbb{E}[dN_t(i)|\mathcal{F}_t]=\langle \theta(i)\rangle_t\,dt, \quad
i\not=i^*.\] 
That is the number of jumps in the direction $i$ in the time interval $[t,
t+dt)$ depends on the past of the process and is equal to $\langle
\theta(i)\rangle_t\,dt$ in mean. We may go a little further and compute the
generating function of those jumps. Indeed, since the conditional measure
$\mathbb{E}[\cdot|\mathcal{F}_t]$ decomposes as a sum,
$\mathbb{E}[\cdot|\mathcal{F}_t]=\sum_\alpha Q_t(\alpha)
\mathbb{E}_\alpha[\cdot|\mathcal{F}_t]$, and since under
$\mathbb{E}_\alpha[\cdot|\mathcal{F}_t]$ the $dN_t(i)$'s are Poisson point
processes with intensity $\theta(i|\alpha)dt$, we have
\begin{eqnarray}\label{gener-poiss}
\log \mathbb{E}[e^{z\,dN_t(i)}|\mathcal{F}_t]= dt\,\langle \theta(i)\rangle_t\,(e^z-1),
\end{eqnarray}
with $\langle \theta(i)\rangle_t=\sum_\alpha Q_t(\alpha) \theta(i|\alpha)$. That
is, under $\mathbb{P}$, the $dN_t(i)$'s are point processes with intensities
$\langle \theta(i)\rangle_tdt$, as announced. As above, a similar computation
shows that the $dN_t(i)$'s, for fixed $t$, are independent variables for
$i\not=j$ under $\mathbb{P}_\alpha$ but not under $\mathbb{P}$. 
An alternative description of this limit is given in ref.\cite{clement}, see also
the forthcoming ref.\cite{BP12}.

\medskip

Up to now, our arguments have been only in law. A rigourous
construction\footnote{Which is an alternative to ref.\cite{clement} in that
  case.} of processes, living on a well-defined probability space, and having
all the required properties, is to deform a suitable a priori measure
$\mathbb{P}^0$. The hint that this is possible is the formula for $Q_t(\alpha)$
obtained by taking the continuous time limit of eq.(\ref{Qsol}). There are some
cancellations of powers of $\delta$ between numerator and denominator yielding
\[Q_t(\alpha)=Q_0(\alpha)\,\frac{ Z_t(\alpha)}{Z_t},\quad {\rm with}\
Z_t:=\sum_\beta Q_0(\beta)Z_t(\beta), \] where
\begin{eqnarray}\label{Zmartin}
  Z_t(\alpha):= \prod_{i\not=i^*}\theta(i|\alpha)^{N_t(i)}\,e^{-t(\theta(i|\alpha)-1)}.
\end{eqnarray}
One recognizes $Z_t(\alpha)$ as the standard exponential Poisson martingale. So,
let us start from an a priori probability measure $\mathbb{P}^0$ accommodating
for independent Poisson processes $N_t(i)$, $i\not=i^*$, of intensity $dt$.
Define $\mathbb{P}_\alpha:=Z_t(\alpha)\,\mathbb{P}^0$ on $\mathcal{F}_t$. Then,
under $\mathbb{P}_\alpha$, the $N_t(i)$'s are independent Poisson processes with
intensity $\theta(i|\alpha)dt$. Defining
\[ \mathbb{P}:= (\sum_\alpha Q_0(\alpha)Z_t(\alpha))\, \mathbb{P}^0,\] it is
plain that the $Q_t(\alpha)$'s are $\mathbb{P}$-martingales and the $N_t(i)$'s
have the law we were after. For instance, since $Q_t(\alpha)=Q_0(\alpha)\,
Z_t(\alpha)/Z_t$, we have $\mathbb{E}[\cdot|\mathcal{F}_t]=\sum_\alpha
Q_t(\alpha) \mathbb{E}_\alpha[\cdot|\mathcal{F}_t]$ so that
$dN_t:=N_{t+dt}(i)-N_t(i)$ is at most $1$ and
$\mathbb{P}[dN_t(i)=1|\mathcal{F}_t]=dt \langle \theta(i)\rangle_t$.

\section{Iterated indirect quantum measurements}
Although purely probabilistic -- involving classical probability only -- the
previous description of iterated stochastic measurements finds applications in
the quantum world, in particular in the framework of repeated indirect
non-demolition measurements \cite{non-demo}. Recall that an indirect quantum
measurement consists in letting a quantum system interact with another
quantum system, called the probe, and implementing a direct Von Neumann
measurement on the probe. One then gains information on the system because the
probe and the system have been entangled. Repeating the cycle of entanglement
and measurement progressively increases the information on the system as in the
model apparatus we described above.

Let $\mathcal{S}$ be the quantum system and $\mathcal{H}_s$ be its Hilbert space
of states. Pick a basis of states $\{|\alpha\rangle\}$ in $\mathcal{H}_s$, which
are going to play the role of pointer states. Let $\mathcal{P}$ be the probe and
$\mathcal{H}_p$ be its Hilbert space.  We assume that the probe-system
interaction preserves the pointer states: a system initially prepared in one of
the pointer state remains in this state after having interacted with the probes.
This requires a peculiar form for the unitary operator $U$ of the probe-system
interaction:
\begin{eqnarray}\label{Upoint}
 U = \sum_\alpha\ket{\alpha}\langle\alpha|\otimes U_\alpha,
\end{eqnarray}
with $U_\alpha$ an unitary operators on $\mathcal{H}_p$. 
Alternatively, $U \ket{\alpha}\otimes\ket{\nu} = \ket{\alpha}\otimes U_\alpha\ket{\nu}$ for any $\ket{\nu}\in\mathcal{H}_p$, a property coding for the fact that the pointer states $\ket{\alpha}$ are preserved by this interaction.

We imagine sending identical copies of the probe, denoted
$\mathcal{P}_1,\mathcal{P}_2,\cdots$, one after the other through the system and
measuring an observable on each probe after the interaction. We assume that the
in-going probes have all been prepared in the same state
$|\psi\rangle\in\mathcal{H}_p$, and that the observables measured in the
out-going channel are all identical with non-degenerate spectrum $I$. Let
$\{\ket{i}\}\in\mathcal{H}_p$, $i\in I$, be the basis of eigenstates of the
measured observable. We denote by $i_k$ the output of the measurement on the
$k^{\rm th}$ out-going probe. In analogy with previous section, we call the
cycle entanglement and measurement on a probe a partial measurement. The results
of repetitions of theses cycles of partial measurements are random sequences
$(i_1,i_2,\cdots)$, $i_k\in I$. As before, such infinite series of partial
measurements will be called a complete measurement. The unitary operator $U$
codes for the probability of measuring a given value $i$ on the out-going probe.
Suppose that the in-going probe has been prepared in the state $\ket{\psi}$ and
the system $\mathcal{S}$ in the state $\ket{\alpha}$.  After interaction, the
system-probe state is $\ket{\alpha}\otimes U_\alpha\ket{\psi}$ and the
probability to measure the value $i$ of the probe observable is
\[ p(i|\alpha):= |\langle i|U_\alpha\ket{\psi}|^2,\] by the rule of quantum
mechanics. So $|\langle i|U_\alpha\ket{\psi}|^2$ is the probability to measure
$i$ in the out-going channel conditioned on the system state be $\ket{\alpha}$.
The analogy with the previous section should start to become clear.

\subsection{Discrete time description}
Let $\rho$ be the system density matrix. The system state probability
distribution is $Q(\alpha)=\langle\alpha|\rho\ket{\alpha}$. The aim of this
section is to describe how the system state distribution and the density matrix evolve when the cycles of entanglement and measurement are repeated, and to make explicit contact with previous sections.

Assume that the system is initially prepared in a density matrix state $\rho_0$,
and let us look at what happens during a cycle of entanglement and interaction.
Recall that the probe is assumed to be prepared in the density matrix state
$\ket{\psi}\langle\psi|$. After interaction, the joint system-probe density
matrix is $U\rho_0\otimes \ket{\psi}\langle\psi|U^{\dag}$. The observable, with
spectrum $I$, is then measured on the probe. If $i_1$ is the output value of
this measurement, the joint system-probe state is projected into $\rho_1\otimes
\ket{i_1}\langle i_1|$ with
\[ \rho_1:= \frac{1}{\pi_0(i_1)}\, \langle i_1|U\ket{\psi}\,\rho_0\, \langle
\psi|U^{\dag}\ket{i_1},\] 
This occurs with probability $\pi_0(i_1)={\rm Tr}\big(\, \langle
i_1|U\ket{\psi}\,\rho_0\, \langle \psi|U^{\dag}\ket{i_1} \,\big)$. Using the
assumed property of $U$, eq.(\ref{Upoint}), this can rewritten as
\[ \pi_0(i):= {\rm Tr}\big(\, \langle i|U\ket{\psi}\,\rho_0\, \langle
\psi|U^{\dag}\ket{i} \,\big)=\sum_\alpha p(i|\alpha)\,Q_0(\alpha) .\] 

How this cycle is to be repeated is clear. Let $\rho_{n-1}$ be the system
density matrix after the $n-1$ first partial measurements -- this density matrix
depends on the random values of these measurements, so that
$\rho_{n-1}=\rho_{n-1}(i_1,\cdots,i_{n-1})$, but we simplify the notation by not
writing explicitly the values of the measurements. We let the system interact
with the $n^{\rm th}$ probe and do a measurement on this probe. If $i_n$
is the output value of this $n^{\rm th}$ partial measurement, the system state
is projected into
\begin{eqnarray}\label{rho_recur} 
\rho_n = \frac{1}{\pi_{n-1}(i_n)}\, \langle i_n|U\ket{\psi}\,\rho_{n-1}\,
\langle \psi|U^{\dag}\ket{i_n}, 
\end{eqnarray}
where again we simplified the notation by not writing the values of the partial
measurements -- $\rho_n$ should have been written as
$\rho_n(i_n|i_1,\cdots,i_{n-1})$ and similarly for $\pi_{n-1}$. This projection
occurs with probability $\pi_{n-1}(i_n)$, with
\[ \pi_{n-1}(i):= {\rm Tr}\big(\, \langle i|U\ket{\psi}\,\rho_{n-1}\, \langle
\psi|U^{\dag}\ket{i} \,\big) =\sum_\alpha p(i|\alpha)\,Q_{n-1}(\alpha) .\] The
diagonal matrix elements of the density matrix are the probabilities for the
system be in a pointer state, that is $Q_n(\alpha)=\langle
\alpha|\rho_n\ket{\alpha}$. From eq.(\ref{rho_recur}) we read that
\[ Q_n(\alpha) = \frac{p(i_n|\alpha)\, Q_{n-1}(\alpha)}{\pi_{n-1}(i_n)}.\]

The two above equations exactly coincide with eqs.(\ref{probpi},\ref{recurr})
defining iterated stochastic measurements. So everything we wrote in the
previous sections applies. In particular the collapse of the system probability
distribution is a discrete implementation of the wave function collapse in Von
Neumann measurement. The quantum system observable measured by the iteration of
cycles of entanglement and indirect measurement is that with eigenstate basis
$\{\ket{\alpha}\}$. The collapse happens only for an infinite sequence of
partial measurement reflecting the fact that the iterated stochastic measurement
apparatus is macroscopic only if an infinite sequence of partial measurements is
implemented, see ref.\cite{qnd_bb}.

\subsection{Continuous time limit}
The aim of this section is to take the continuous time limit of the discrete
recurrence equation (\ref{rho_recur}) for the quantum density matrix using the
results of the previous section. Doing this we will make contact with the
so-called Belavkin equations \cite{Belavkin}, describing continuous time
measurements in quantum mechanics and which are non-linear stochastic
Schr\"odinger equations \cite{StoSchroe}.

The small parameter $\delta$ is the time duration of the system-probe
interaction, so that the unitary operator is $U=\exp(-\imath\delta H)$ with $H$ the
system-probe hamiltonian~\footnote{We use the notation $\imath$, without a dot, to code for the square root of $-1$.}. As is well known, the dynamics of a quantum system
under continuous measurements is frozen by continuous wave packet reductions, a
fact named the quantum Zeno effect. To avoid it, we have to rescale the
system-probe interaction at the same time we decrease the interaction time
duration. So we assume the following form of the hamiltonian $H$,
\begin{eqnarray}\label{Hdecomp}
 H= H_s\otimes 1 + 1\otimes H_p + \frac{1}{\sqrt{\delta}}\, H_I, 
 \end{eqnarray}
where $H_s$ is the system hamiltonian, $H_p$ the probe hamiltonian and $H_I$ the
interaction hamiltonian.

For the pointer state to be stable under the action of $U=e^{-\imath\delta H}$, eq.(\ref{Upoint}),
we should assume that $H_s$ is diagonal in the pointer basis, $H_s=\sum_\alpha |\alpha\rangle E_\alpha \langle\alpha|$ for some energies $E_\alpha$, -- this is linked to the non-demolition character of the measurement -- and that 
\[ H_I=\sum_\alpha \ket{\alpha}\langle\alpha|\otimes H_\alpha, \]
with $H_\alpha$ acting on $\mathcal{H}_p$ but $\alpha$ dependent. 
The conditioned probabilities
$p(i|\alpha)$ are then
\[ p(i|\alpha)= |\langle i\ket{\psi} - \imath\sqrt{\delta}\, \langle
i|H_\alpha\ket{\psi}+\cdots|^2,\] 
so that the Brownian diffusive case corresponds $\langle i\ket{\psi}\not=0$ and
the Poissonian jumpy case to $\langle i\ket{\psi}=0$.

In both cases, the continuous time limit is then obtained by performing the
scaling limit $\delta\to0$, $n\to\infty$ with $t:=n/\delta$ fixed as above.

It is useful to recast the quantum recursion relation (\ref{rho_recur}) into a
difference equation. This simplifies matter when taking the continuous time
limit. Let us write $\rho_n=\sum_i \rho_{n}(i_n)\, \mathbb{I}_{i_n=i}$ with
$\rho_{n}(i_n)$ defined in eq.(\ref{rho_recur}). Recall that
$\mathbb{E}[\mathbb{I}_{i_n=i}|\mathcal{F}_{n-1}]=\pi_{n-1}(i)$ and write
$\mathbb{I}_{i_n=i}= (\mathbb{I}_{i_n=i}-\pi_{n-1}(i))+\pi_{n-1}(i)$. This leads
to the Doob decomposition of the difference $\rho_n-\rho_{n-1}$ as,
\begin{eqnarray}\label{rho_diff} 
\rho_n-\rho_{n-1}= (D\rho)_{n-1} + (\Delta\rho)_{n},
\end{eqnarray}
with $(D\rho)_{n-1}:=\mathbb{E}[\rho_n|\mathcal{F}_{n-1}]-\rho_{n-1}$, which is
$\mathcal{F}_{n-1}$-measurable, and
$(\Delta\rho)_{n}:=\rho_n-\mathbb{E}[\rho_n|\mathcal{F}_{n-1}]$, which satisfies
$\mathbb{E}[(\Delta\rho)_{n}|\mathcal{F}_{n-1}]=0$. Explicitely,
\begin{eqnarray*}
  (D\rho)_{n-1}&=& \sum_i \langle i|U\ket{\psi}\,\rho_{n-1}\, \langle
  \psi|U^{\dag}\ket{i} -\rho_{n-1},\\ 
  (\Delta\rho)_{n}&=& \sum_i \frac{\langle i|U\ket{\psi}\,\rho_{n-1}\, \langle
    \psi|U^{\dag}\ket{i}}{\pi_{n-1}(i)}\, (X_n(i)-X_{n-1}(i)), 
\end{eqnarray*}
where we used $\mathbb{I}_{i_n=i}-\pi_{n-1}(i)=X_n(i)-X_{n-1}(i)$, as in
previous section. In the continuous time limit, the first term $(D\rho)_{n-1}$
is going to converge towards the drift term and the second one
$(\Delta\rho)_{n}$ to the noisy source of the stochastic differential equation.

\subsubsection{Brownian diffusive limit}

The Brownian diffusive limit occurs when $\langle i\ket{\psi}\not=0$ for all
$i$. Then $p(i|\alpha)\simeq p_0(i)( 1+\sqrt{\delta}\, \Gamma(i|\alpha)+\cdots)$
for $\delta$ small with,
\[ p_0(i)=|\langle i\ket{\psi}|^2,\quad \Gamma(i|\alpha)=2{\rm Im}\,
\big(\frac{\langle i|H_\alpha\ket{\psi}}{\langle i\ket{\psi}}\big).\]
This is the situation we encountered in the previous section on the classical
diffusive limit, so that we can borrow all results obtained there.

It is then a simple matter to naively take the continuous time limit of the
difference equations (\ref{rho_diff}). This limit exists only if $\langle\psi|H_I\ket{\psi}=0$, 
which is equivalent to
\[ \langle\psi|H_\alpha\ket{\psi}=0\ {\rm for\ all}\ \alpha,\]
a criteria which we assume to hold true. Recall
that this scaling limit consists in $\delta\to0$ with $t=n\delta$ fixed. 
Let us first expand the term $(D\rho)_{n-1}$ in power of $\sqrt{\delta}$. The term of order $\sqrt{\delta}$ vanishes due to the condition $\langle\psi|H_I\ket{\psi}=0$, and for the term of order $\delta$ we get: $(D\rho)_{[t/\delta]}
\simeq_{\delta\to0}L_d(\rho_{t})\delta$, with Linbladian
\[ L_d(\rho_{t}):=-\imath[H_{s},\rho_t]+\sum_{i}p_{0}(i)(C_{i}\,\rho_t\,
C_{i}^{\dagger}-\frac{1}{2}\{C_{i}^{\dagger}C_{i},\rho_t\}),\]
where we defined the operators $C_i$'s acting on $\mathcal{H}_s$ by $C_i:=-\imath\frac{\langle i\vert H_{I}\vert\psi\rangle}{\langle i\vert\psi\rangle}$, or equivalently 
\[C_{i}:=-\imath\sum_\alpha \ket{\alpha} \frac{\langle i\vert H_{\alpha}\vert\psi\rangle}{\langle i\vert\psi\rangle}\langle\alpha|,\]
using the decomposition of $H_I$ on pointer states.
Remark that $\sum_ip_0(i)C_i=0$ thanks to the assumed condition $\langle\psi|H_\alpha\ket{\psi}=0$.
Similarly, expanding the  term $(\Delta \rho)_n$ using $\pi_{n-1}(i)\simeq p_0(i)[1+\delta {\rm
  Tr}[(C_{j}+C_{j}^{\dagger})\rho_{t}]+\cdots]$, we get
$\lim_{\delta\to0}(\Delta\rho)_{[t/\delta]}=\sum_j \mathcal{D}_j(\rho_t)\,
dX_{t}(j)$, with
\[ \mathcal{D}_j(\rho_t):=C_{j}\rho_{t}+\rho_{t}C_{j}^{\dagger}-\rho_{t}{\rm
  Tr}[(C_{j}+C_{j}^{\dagger})\rho_{t}].\] 
Note that computing these limits only uses the decomposition (\ref{Hdecomp}) of the hamiltonian $H$
and not the existence of a preferred pointer state basis\footnote{The existence
  of the pointer state basis was however used in determining the statistical
  properties of the fields $X_t(i)$.}. Gathering shows that the Brownian limit
of the difference equation (\ref{rho_recur}) is
\begin{eqnarray}\label{belav_diffu}
d\rho_t= L_d(\rho_t)\, dt + \sum_j \mathcal{D}_j(\rho_t)\, dX_{t}(j)
\end{eqnarray}
where the $X_t(j)$'s are the Gaussian centred processes, with quadratic variation
\[ dX_t(i)dX_t(j)= (p_0(i)\delta_{i;j}-p_0(i)p_0(j))\, dt,\]
defined in eq.(\ref{brown}) and in the discussion around this equation. 
This is an example of the diffusive Belavkin equation \cite{Belavkin,Barchielli2009}. 
It is important to recall that $\sum_i p_0(i) C_i=0$ since without this condition, but with $\sum_i X_t(i)=0$ as we do have, eq.(\ref{belav_diffu}) would not be positivity preserving \cite{Barchielli2009}. Contact with previous sections can be made explicit by recalling that the state probability distribution is $Q_t(\alpha)=\langle\alpha|\rho_t\ket{\alpha}$ and by noticing that ${\rm
  Tr}[(C_{j}+C_{j}^{\dagger})\rho_{t}]=\langle\Gamma(i)\rangle_t$.
We only took a naive limit of the difference equation (\ref{rho_recur}), to mathematically
prove the diffusive Belavkin equation for the system density matrix in the
scaling limit would require more delicate arguments.

\subsubsection{Poissonian jumpy limit}
The Poissonian limit occurs when $\langle i\ket{\psi}=0$. This cannot happen for
all $i$ as $\{\ket{i}\}$ forms an orthonormal basis of $\mathcal{H}_p$ and
$\ket{\psi}$ is non zero. So, we assume, for simplicity, that one element of
this basis is $\ket{\psi}$, say $\ket{i^*}=\ket{\psi}$, and all others are
orthogonal to $\ket{\psi}$, i.e. $\langle i\ket{\psi}=0$ for all $i\not=i^*$.
Then, $p_0(i^*|\alpha)\simeq_{\delta\to0}1$ and
$p_0(i|\alpha)\simeq_{\delta\to0} \delta\, \theta(i|\alpha)$, for $i\not=i^*$,
with
\[ \theta(i|\alpha)= |\langle i |H_\alpha \ket{\psi}|^2.\]
This is the situation we encountered in the previous section on the classical
Poisson jumpy limit, so that we can borrow all results obtained there.

As in the diffusive case, it is a simple matter to naively take the continuous
time limit of the difference equation (\ref{rho_diff}). This only uses the
decomposition the hamiltonian $H$ but the limit exists only if
$\langle\psi|H_I\ket{\psi}=0$, and we assume this to be true. Expanding the first
term $(D\rho)_{n-1}$ to second order in $\sqrt{\delta}$, we get $(D\rho)_{[t/\delta]} \simeq_{\delta\to0} L_p(\rho_{t})\delta$,
with Linbladian
\[L_p(\rho_{t}):=-\imath[H_{s},\rho_t]+\sum_{i\not=i^*}(D_{i}\,\rho_t\,D_{i}^{\dagger}
-\frac{1}{2}\{D_{i}^{\dagger}D_{i},\rho_t\}),\] 
where we defined the operators $D_{i}:=-\imath\, {\langle i\vert H_{I}\vert\psi\rangle}$
acting on $\mathcal{H}_s$, that is
\[ D_i:= -\imath\,\sum_\alpha \ket{\alpha} \langle i\vert H_{\alpha}\vert\psi\rangle \langle\alpha| .\]
To compute the limit of the second term $(\Delta\rho)_n$, we notice that, to leading order in $\delta$, $\langle i|U\ket{\psi}\rho\langle\psi| U \ket{i}\simeq \delta\, D_i\rho D_i^\dag$ and $\pi_{n-1}(i)\simeq \delta\,  {\rm Tr}[D_{i}\rho_{t}D_{i}^{\dagger}]$ for $i\not= i^*$, and we get
$\lim_{\delta\to0}(\Delta\rho)_{[t/\delta]}=\sum_{i\not=i^*} \widehat{\mathcal{D}}_i(\rho_t)\, dY_{t}(i)$, with
\[ \widehat{\mathcal{D}}_i(\rho_t):=\frac{D_{i}\,\rho_{t}\,D_{i}^{\dagger}}{{\rm Tr}[D_{i}\rho_{t}D_{i}^{\dagger}]}-\rho_t,\] 
where the last term $-\rho_t$ comes from using $dY_t(i^*)=-\sum_{i\not=i^*}dY_{t}(i)$ when computing the contribution of the $i^*$-term in $(\Delta\rho)_n$, as in previous section.
Gathering shows that the Poissonian limit of the difference equation
(\ref{rho_recur}) is
\begin{eqnarray}\label{belav_poiss}
d\rho_t= L_p(\rho_t)\, dt +\sum_{i\not=i^*} \widehat{\mathcal{D}}_i(\rho_t)\,
dY_{t}(i) 
\end{eqnarray}
where the $Y_t(j)$'s are the Poisson-like compensated martingales defined in
eq.(\ref{dpoisson}) above. That is,
\[dN_t(i)= dY_t(i) + {\rm Tr}[D_{i}\rho_{t}D_{i}^{\dagger}]\,dt, \quad i\not=i^*,\]
where $dN_t(i)$'s are the point processes with intensities
$\langle\theta(i)\rangle_t\,dt$ defined in previous section, see e.g.
eq.(\ref{gener-poiss},\ref{Zmartin}). Note that, using the decomposition of the
interaction hamiltonian $H_I$ on the pointer state basis, $H_I=\sum_\alpha
\ket{\alpha}\langle\alpha|\otimes H_\alpha$, we have
\[ {\rm Tr}[D_{i}\rho_{t}D_{i}^{\dagger}] = \sum_\alpha \theta(i|\alpha)\,
Q_t(\alpha)= \langle\theta(i)\rangle_t,\] so that the previous equation indeed
coincides with eq.(\ref{dpoisson}). Equation (\ref{belav_poiss}) is an example
of a jumpy Belavkin equation \cite{Belavkin}. 
Let us finally point out that the stochastic processes (\ref{belav_diffu},\ref{belav_poiss}) are not the most general one because we assumed that they preserve the pointer state basis~\footnote{We made this assumption when computing the probability distributions of the processes $X_t(i)$ and $Y_t(i)$ but not when formally taking the continuous time limit of the discrete eq.(\ref{rho_recur}).} so that the operator $H_s$, $C_i$ or $D_j$ are diagonal in the pointer basis. This is of course related to the non-demolition property. Eqs.(\ref{belav_diffu},\ref{belav_poiss}) are also peculiar examples of more general class of models for continuous quantum measurements whose long time behavior leads to purification of mixed states, see e.g. \cite{Barchielli2003,Barchielli2009}.

\bigskip

{\it Acknowledgements}: This work was in part supported by ANR contract ANR-2010-BLANC-0414.01 and ANR-2010-BLANC-0414.02.

%\end{references}

\end{document}